\documentclass[aps,prd,preprint,nofootinbib]{revtex4-1}
\usepackage{amsmath,amssymb,graphics,graphicx,color}
\usepackage{scrextend}
\usepackage{hyperref}
\usepackage{float}
\usepackage{xcolor}
\usepackage[utf8]{inputenc}

\usepackage{multirow}
\usepackage{bigcenter}
\usepackage{caption}
\usepackage{subcaption}

\begin{document}

\title{Axion Like Particle Search at Higgs Factories }
\author{Kingman Cheung,$^{a,b,c}$ C.J. Ouseph$^{a,b}$}
\affiliation{
  $^a$ Department of Physics, National Tsing Hua University, Hsinchu 30013,
  Taiwan\\
  $^b$ Center for Theory and Computation, National Tsing Hua University, Hsinchu 30013, Taiwan \\
  $^c$ Division of Quantum Phases and Devices, School of Physics,
  Konkuk University, Seoul 143-701, Republic of Korea}

\date{\today}
\begin{abstract}
We study the potential of the future Higgs factories, including the ILC,
CEPC, and FCC-ee with $\sqrt{s}$ = 240-250 GeV on discovering
axion-like particles (ALPs) through various production channels in 
the leptonic final states, $e^+e^- \to f\bar{f} a$, 
where $f=e,\mu,\nu$.
We show that the $e^+e^- \to e^+e^- a$ with $a \to \gamma\gamma$ 
provides the best bounds for the $g_{a\gamma\gamma}$ and $g_{aZZ}$ couplings,
while $e^+e^- \to \nu\bar{\nu}a$, with $a \to \gamma\gamma$ offers 
the best bounds for the $g_{aZZ}$ and $g_{aZ\gamma}$ couplings. 
The $e^+e^- \to \mu^+\mu^- a$ with $ a \to \gamma\gamma$ 
provides intermediate sensitivity to the $g_{aZZ}$ coupling.
Our estimates of the bounds for the $g_{a\gamma\gamma}$, $g_{aZ\gamma}$,
and $g_{aZZ}$ couplings as a function of ALP mass ($M_a$) ranging 
from 0.1 GeV to 100 GeV provide valuable insights for future experiments 
aiming to detect ALPs. 
We find that $g_{a\gamma\gamma}$ around $1.5\times10^{-4}~\rm GeV^{-1}$ for $M_a = 0.1-6$ GeV is currently not ruled out
by any other experiments. 
 \end{abstract}

\maketitle
\section{Introduction}
The strong CP problem in the standard model (SM) is a long-standing problem \cite{Peccei:1977hh}. The best solution comes by introducing a global $U(1)_{PQ}$ symmetry, which was spontaneously broken down by a dynamical axion field. The resulting pseudo-Nambu-Goldstone boson is known as the QCD axion \cite{Peccei:1977hh,Weinberg:1977ma,Wilczek:1977pj}. It can also serve as a dark matter candidate \cite{Preskill:1982cy,Abbott:1982af,Dine:1982ah}.

Nonobservation of the neutron electric dipole moment demands the breaking scale of the PQ symmetry to be very high with $f_a > 10^9$ GeV, implying a tiny mass to the axion and very small couplings to the SM particles. If we do not require the pseudo-Nambu-Goldstone boson to be the  solution of the strong CP problem, the mass of the axion is  not restricted by the breaking scale $f_a$. Such a hypothetical particle, coined as axion-like particle (ALP), is also a pseudoscalar boson. 

However, the ALP remains one of the possible dark matter candidates.
The axion mass and couplings to SM particles can extend over
many orders of magnitude, which are only constrained by 
astrophysical and cosmological observations, 
as well as collider experiments. In this work, we consider the potential sensitivities on the
parameter space of the ALP model by 
searching for such ALPs in the proposed Higgs factories, including 
the International Linear Collider (ILC) \cite{ILC:2007bjz}, 
CEPC \cite{CEPC-SPPCStudyGroup:2015csa}, and 
FCC-ee \cite{FCC:2018evy} with $\sqrt{s} = 240-250$ GeV.
We consider the following leptonic production channels
$e^+ e^- \to f \bar f a$ with $f=e,\mu,\nu $. Given the
center-of-mass energy is only 250 GeV, we consider the ALP mass 
in the range of $0.1- 100$ GeV.
Typical Feynman diagrams for production can be found in Fig.~\ref{fig.1}.

We focus on the diphoton decay mode of the ALP, which is shown to be
dominant.  Thus, we have rather clean final states 
$ f\bar f (\gamma\gamma)$ with $f=e,\mu,\nu$. The SM background is 
calculated and found to be small. Finally, we show the sensitive regions
of the couplings.
The organization of this work is as follows. In the next section, we 
describe the model and existing constraints. In Sec. III, we show the
signal-background analysis. We calculate the sensitivities of the
ALP couplings in Sec. IV. We summarize in Sec. V.

\section{Theoretical Setup}
\subsection{Model}
The axion, as a pseudo-Goldstone boson, has derivative couplings to fermions, as well
as $CP$-odd couplings to the gauge field strengths. Before rotating 
the $B$ and $W^i$ fields to
the physical $\gamma,~Z,~W^\pm$, the interactions of the 
axion are given by \cite{Brivio:2017ije,Georgi:1986df,Ren:2021prq}
\begin{equation}\label{Eq.1}
\mathcal{L}=\mathcal{L}_f+\mathcal{L}_{g}+\mathcal{L}_{BB}+\mathcal{L}_{WW}
\end{equation}
where,
\begin{equation*}
    \mathcal{L}_f=-\frac{ia}{f_a}\sum_{f}g_{af}~m^{diag}_f\bar{f}\gamma_5f
\end{equation*}
\begin{equation*}
    \mathcal{L}_g = -C_g\frac{a}{f_a}G^A_{\mu\nu}\Tilde{G}^{\mu\nu,A}
\end{equation*}
\begin{equation*}
    \mathcal{L}_{BB} = -C_{BB}\frac{a}{f_a}B_{\mu\nu}\Tilde{B}^{\mu\nu}
\end{equation*}
\begin{equation*}
    \mathcal{L}_{WW} = -C_{WW}\frac{a}{f_a}W^i_{\mu\nu}\Tilde{W}^{\mu\nu,i}.
\end{equation*}
where $a$ represents the ALP field, $f_a$ is the ALP decay constant, $A=1,....8$ is the $SU(3)$ color index and $i=1,2,3$ is the $SU(2)$ index. The $B,W^3$ fields rotated into $\gamma, Z$ by
\begin{equation}\label{Eq.2}
\begin{pmatrix}
W^3_\mu \\
B_\mu 
\end{pmatrix}= \begin{pmatrix}
c_w & s_w \\
-s_w & c_w
\end{pmatrix} \begin{pmatrix}
Z_\mu \\
A_\mu 
\end{pmatrix}.
\end{equation}

where $c_w$,$s_w$ are cosine and sine of the Weinberg angle. 
The axion interactions with the fermion and the physical gauge bosons are given by
\begin{equation}\label{Eq.3}
\begin{split}
\mathcal{L}=-\frac{ia}{f_a}\sum_{f}g_{af}m^{diag}_f\bar{f}\gamma_5f-C_g\frac{a}{f_a}G_{\mu\nu}^A\tilde{G}^{\mu\nu A}-\frac{a}{f_a}\big[(C_{BB}c^2_w+C_{WW}s^2_w)F_{\mu\nu}\tilde{F}_{\mu\nu}+&\\(C_{BB}s^2_w+C_{WW}c^2_w)Z_{\mu\nu}\tilde{Z}_{\mu\nu}+2(C_{WW}-C_{BB})c_ws_wF_{\mu\nu}\tilde{Z}_{\mu\nu}+C_{WW} W^+_{\mu\nu} \tilde{W}^{- \mu\nu}\big]
\end{split}
\end{equation}
The dimensionful couplings associated with ALP interactions from \ref{Eq.3} is given by;
\begin{equation}\label{Eq.4}
g_{a\gamma\gamma}=\frac{4}{f_a}(C_{BB}c_w^2+C_{WW}s_w^2),
\end{equation}
\begin{equation}\label{Eq.5}
g_{aWW}=\frac{4}{f_a}C_{WW},
\end{equation}
\begin{equation}\label{Eq.6}
g_{aZZ}=\frac{4}{f_a}(C_{BB}s_w^2+C_{WW}c_w^2),
\end{equation}
\begin{equation}\label{Eq.7}
g_{aZ\gamma }=\frac{8}{f_a}s_wc_w(C_{WW}-C_{BB}) \,.
\end{equation}

\subsection{Existing Constraints on ALPs }
The experimental bounds on the couplings of ALP to gluons, photons, 
weak gauge bosons, and fermions have been thoroughly investigated 
in numerous sources 
\cite{ParticleDataGroup:2014cgo,Vinyoles:2015aba,Raffelt:2006cw,Friedland:2012hj,Ayala:2014pea,CMS:2014jvv,ATLAS:2015qlt,XENON100:2014csq,Viaux:2013lha}, including their effects at colliders when $f_a$ is approximately at 
the TeV scale \cite{ATLAS:2015qlt, Jaeckel:2015jla}.
Moreover, more recent works had constrained the coupling of ALPs to 
the $W\pm$ gauge boson can be found in 
Ref.~\cite{Dolan:2014ska,Izaguirre:2016dfi}.

\begin{itemize}
    \item  The LEP and the current LHC experiments 
    can probe a significant region of parameter space for ALPs with  
    mass $M_a\geq 5 GeV$. The LEP utilized the 
    $e^+e^-\to\gamma a, \;\; (a\to\gamma\gamma)$ and
    $Z\to a\gamma$ processes \cite{Jaeckel:2015jla} to search for
    ALPs, while ATLAS amd CMS employed the process
    $\gamma\gamma \to a \to \gamma\gamma$ 
    in $\rm PbPb$ collisions at the LHC \cite{dEnterria:2021ljz}. 
    In addition, the rare decay of the Higgs boson 
    $h\to Z a, \; (a\to\gamma\gamma)$ and 
    $h\to a a \to (\gamma\gamma) (\gamma\gamma)$ at the LHC \cite{Bauer:2017nlg} had been utilized to explore the ALP-photon coupling $g_{a\gamma\gamma}$ in relation to the ALP mass $M_a$.
    
	\item The ALPs with masses below the MeV scale has been 
    extensively studied in cosmological and astrophysical observations,
    which have resulted in numerous constraints on ALP couplings,
    including BBN, CMB, and Supernova 1987A \cite{Raffelt:2006cw,Marsh:2015xka}. 
    Furthermore, light ALPs can potentially become the cold dark matter 
    \cite{Preskill:1982cy,Abbott:1982af,Dine:1982ah}, which could lead 
    to their detection through various astrophysical and terrestrial
    anomalies \cite{Arias:2012az}, such as the unexpected X-ray emission
    line at around 3.5 keV \cite{Jaeckel:2014qea} and the excess of 
    electronic recoil events in XENON1T \cite{Gao:2020wer}. 
    These results demonstrated the importance of further exploration and
    investigation into the properties and behavior of ALPs.
    
    \item In the mass range of MeV to GeV, ALPs can significantly impact
    low-energy observables in particle physics. Recent studies in the
    intensity frontier \cite{Izaguirre:2016dfi,Dolan:2017osp,Bauer:2019gfk,NA64:2020qwq} 
    have explored numerous potential search avenues. Examples include
    lepton-flavor-violating decays \cite{Calibbi:2020jvd}, 
    rare meson decays \cite{Izaguirre:2016dfi,Dolan:2017osp,Bjorkeroth:2018dzu,Chakraborty:2021wda}, 
    and ALP production in beam dump experiments \cite{Dobrich:2015jyk}.
    Furthermore, this range of ALPs has been proposed as a possible explanation for the muon anomalous magnetic moment \cite{Bauer:2019gfk,Marciano:2016yhf} 
    and may also provide a feasible solution to the Koto anomaly \cite{Gori:2020xvq}. 
    These findings highlight the importance of continued research into 
    ALPs and their potential implications in particle physics.
    
    \item The search for the process $e^+e^- \to \gamma a$ with
    $ a \to \gamma\gamma$ has recently been conducted by Belle II
    \cite{Belle-II:2020jti} for the ALP mass ranging between 0.1 GeV 
    and 10 GeV. The data utilized in this search corresponded to an integrated luminosity of $(445\pm3)pb^{-1}$, and the mass range explored was 0.2 GeV $<M_a<$ 9.7 GeV.
    
    \item The process $e^+e^- \to e^+e^- a$ with $a \to \gamma\gamma$ 
    at ILC has recently been studied in Ref.~\cite{Steinberg:2021iay,Steinberg:2021wbs,Yue:2021iiu}. 
    Ref.~\cite{Steinberg:2021wbs} showed that the ILC running at 
    $\sqrt{s} = 250$ GeV or $\sqrt{s} = 500$ GeV can discover ALPs in this range of masses with significantly smaller couplings to the SM than previous experiments, down to $g_{aBB}$ = $10^{-3}~\rm TeV^{-1}$. 
    Ref.~\cite{Steinberg:2021iay} showed that with more than $10^9$ 
    $Z$ bosons produced in the Giga-Z mode of the future ILC experiment 
    equipped with the high granular nature of the detector, one 
    can discover of the ALPs coupled to hypercharge with couplings down to nearly $10^{-5}~\rm GeV^{-1}$ over the mass range from 0.4 to 50 GeV.
\end{itemize}

\begin{table}[t!]
 \begin{ruledtabular}
  \begin{tabular}{ccc}
      $e^+ e^-$ Collider &  $\sqrt{s}$  (GeV) &  
     Integrated Luminosity (fb$^{-1}$) \\
       \hline 
       ILC       &  250  & 2000 \\
       CEPC      &  240  & 5600 \\
       FCC-ee    &  250  &  5000 
    \end{tabular}
    \end{ruledtabular}
      \caption{\label{eecollider} 
      A few proposals of $e^+ e^-$ colliders running as a Higgs factory,
      at which the center-of-mass energy and integrated luminosity 
      are shown.
      }
\end{table}   
   
A few proposals of Higgs factories are put forward, including
the ILC \cite{ILC:2007bjz}, CEPC \cite{CEPC-SPPCStudyGroup:2015csa} , 
and FCC-ee \cite{FCC:2018evy}, running at center-of-energies at 
$\sqrt{s} = 240 -250$ GeV with the nominal luminosities shown in 
Table~\ref{eecollider}.
One of the main goals is to carry out
the precision study of the Higgs boson couplings.
We investigate the potential search for ALPs
$e^+e^-$ collisions at the Higgs factories.  Without loss of
generality, we choose $\sqrt{s} = 250$ GeV and a conservative integrated 
luminosity of 2 ab$^{-1}$.

\begin{figure}[th!]
	\includegraphics[width=16cm,height=9cm]{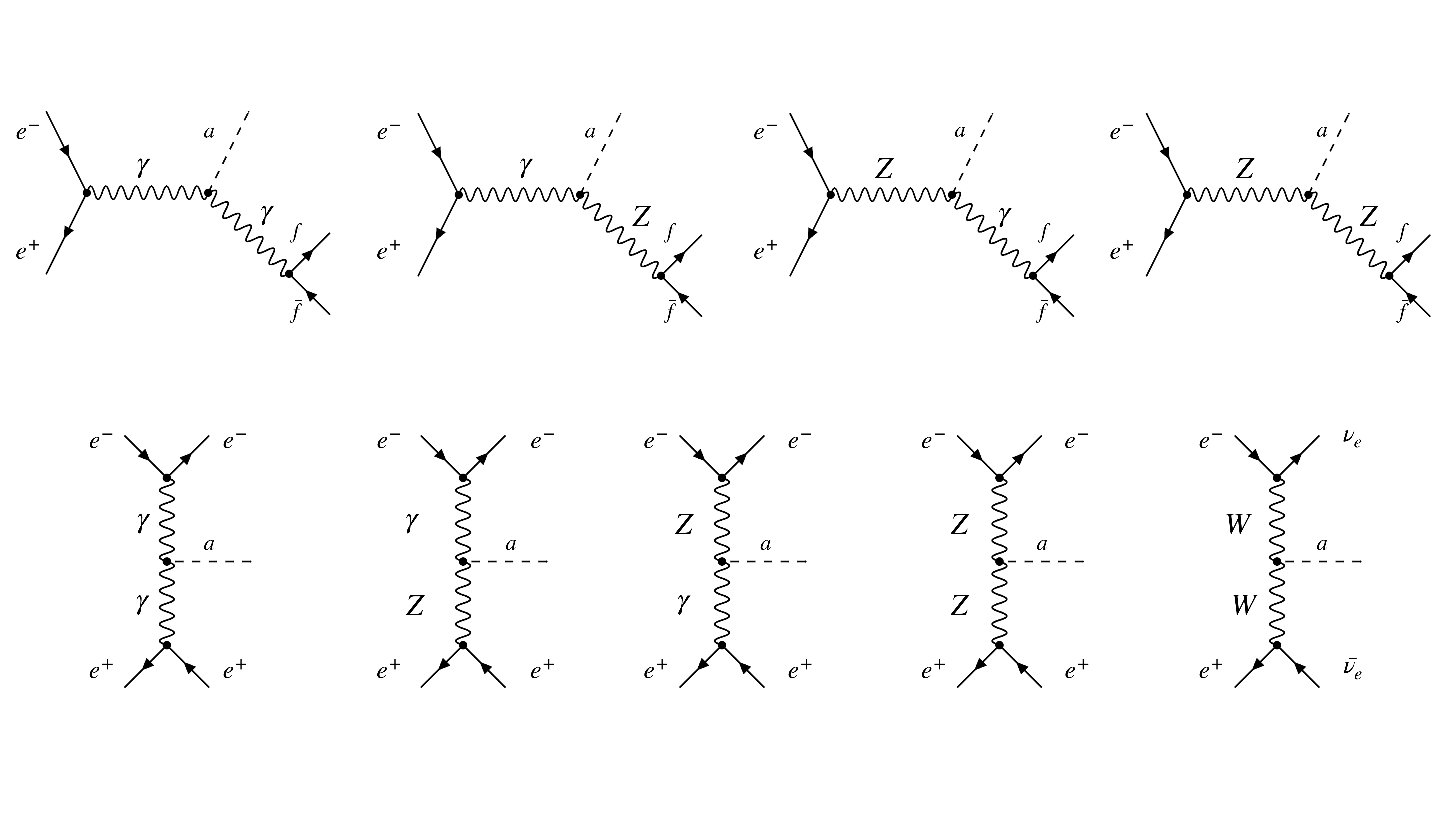}
	\caption{ \small\label{fig.1} 
 Typical Feynman diagrams for production of axion-like particles $a$
 via the process $e^+ e^- \to f \bar f a$ at $e^+ e^-$ collisions, 
 where $f=e,\mu,\nu$. }
\end{figure}

At the Higgs factories, the leptonic processes that we consider are
$e^+e^- \to f\bar{f}a$ where $f = e$, $\mu$, or $\nu$, followed by
$a \to \gamma\gamma$.
This study explores the effects of the coupling
$g_{a\gamma\gamma},\,g_{a \gamma Z},\, g_{a ZZ},\,g_{aWW}$
on the production rates of the ALP.
Typical contributing Feynman diagrams are shown in Fig.~\ref{fig.1}.
Among the diagrams, there are $s$- and $t$-channel diagrams with 
the ALP bremsstrahlung off an internal $\gamma,Z$, or $W$ propagator. 

\section{Signal versus Background }
We use $\rm MadGraph5aMC@NLO$ \cite{Alwall:2011uj,Alwall:2014hca} to 
generate events for the production of ALPs at $e^+ e^-$ collisions. We consider the following channels for detecting the ALP signal:
\begin{itemize}
    \item $e^+ e^- \to~e^+e^- a$ with $a\to \gamma\gamma$\\
    To obtain the production cross-sections of the ALP with mass 
    from $M_a=0.1~\rm GeV$ to $100~\rm GeV$, we apply the following
    initial cuts on the transverse momentum $p_{T}^e$ and rapidity 
    $|\eta^e|$ of the electrons in the final state, as well as the transverse momentum $p_{T}^\gamma$ and rapidity $|\eta^{\gamma}|$ of the photons in final state.
    \begin{itemize}
        \item $ {p_T^{e}}_{\rm min}=10~\rm GeV$
        \item $|\eta^e_{\rm max}|= 1.83$~~($|\cos\theta_{e}|<0.95$)
        \item ${p_T^{\gamma}}_{\rm min}=10~\rm GeV$
        \item $|\eta^{\gamma}_{\rm max}|=2.5$
    \end{itemize}
    
    \item $e^+e^-\to~\mu^+\mu^- a$ with $a\to \gamma\gamma$\\
The final state consisting of muons ($\mu\pm$) and a pair of photons from the ALP decay are selected using the same cuts as the electron case
    \begin{itemize}
        \item ${p_T^{\mu}}_{\rm min}=10~\rm GeV$
        \item $|\eta^{\mu}_{\rm max}|=1.83$~~
        ($|\cos\theta_{\mu}|<0.95$)
        \item ${p_T^{\gamma}}_{\rm min}=10~ \rm GeV$
        \item $|\eta^{\gamma}_{\rm max}|=2.5$
    \end{itemize}
    \item $e+e^-\to \nu\bar{\nu}a$ with $a\to \gamma\gamma$\\
    Here the ALP is produced along with neutrinos in the final states are 
    selected using the following cuts on the rapidity and transverse momentum of the photon and missing transverse energy of $E_T$.
    \begin{itemize}
        \item $\not\!{/}E_T^{\rm min}=20~\rm GeV$
        \item ${p_T^{\gamma}}_{\rm min}=10~\rm GeV$
        \item $|\eta^{\gamma}_{\rm max}|=2.5$       
    \end{itemize}
\end{itemize}
The corresponding irreducible background is also subject to the same cuts 
as discussed above. 
We use $C_{WW}=2$, $C_{BB}=1$, and $f_a = 1$ TeV in calculating the
ALP cross sections, so the corresponding 
coupling strengths $g_{a\gamma\gamma}$, $g_{a Z\gamma }$, $g_{aZZ}$,
and $g_{aWW}$ are obtained using Eqs.~(\ref{Eq.4}) -- (\ref{Eq.7}) and
listed in Table~\ref{Tb.1}.
Note that we have chosen different values for $C_{WW}$ and $C_{BB}$,
otherwise $g_{aZ\gamma}$ would vanish.

\begin{table}[th!]
\caption{\label{Tb.1}
The ALP coupling strengths $g_{a\gamma\gamma}$, $g_{aZ\gamma}$,
$g_{aZZ}$. and $g_{aWW}$ with 
$C_{WW}=2,~C_{BB}=1,~f_a=10^{3}~\rm GeV$ using 
Eqs.~(\ref{Eq.4}) -- (\ref{Eq.7}).
}
\begin{ruledtabular}
\begin{tabular}{cc}
ALP couplings & Numerical Value ($\rm GeV^{-1}$)~ \\
\hline
$g_{a\gamma\gamma}$ & $4.88\times10^{-3}$\\
\hline
$g_{aZ\gamma}$ & $1.38\times10^{-3}$  \\
\hline
$g_{aZZ}$ & $7.11\times10^{-3}$  \\
\hline
$g_{aWW}$ & $8\times10^{-3}$ 
\end{tabular}
\end{ruledtabular}
\end{table}

\begin{figure}[th!]
	\includegraphics[width=13cm,height=9cm]{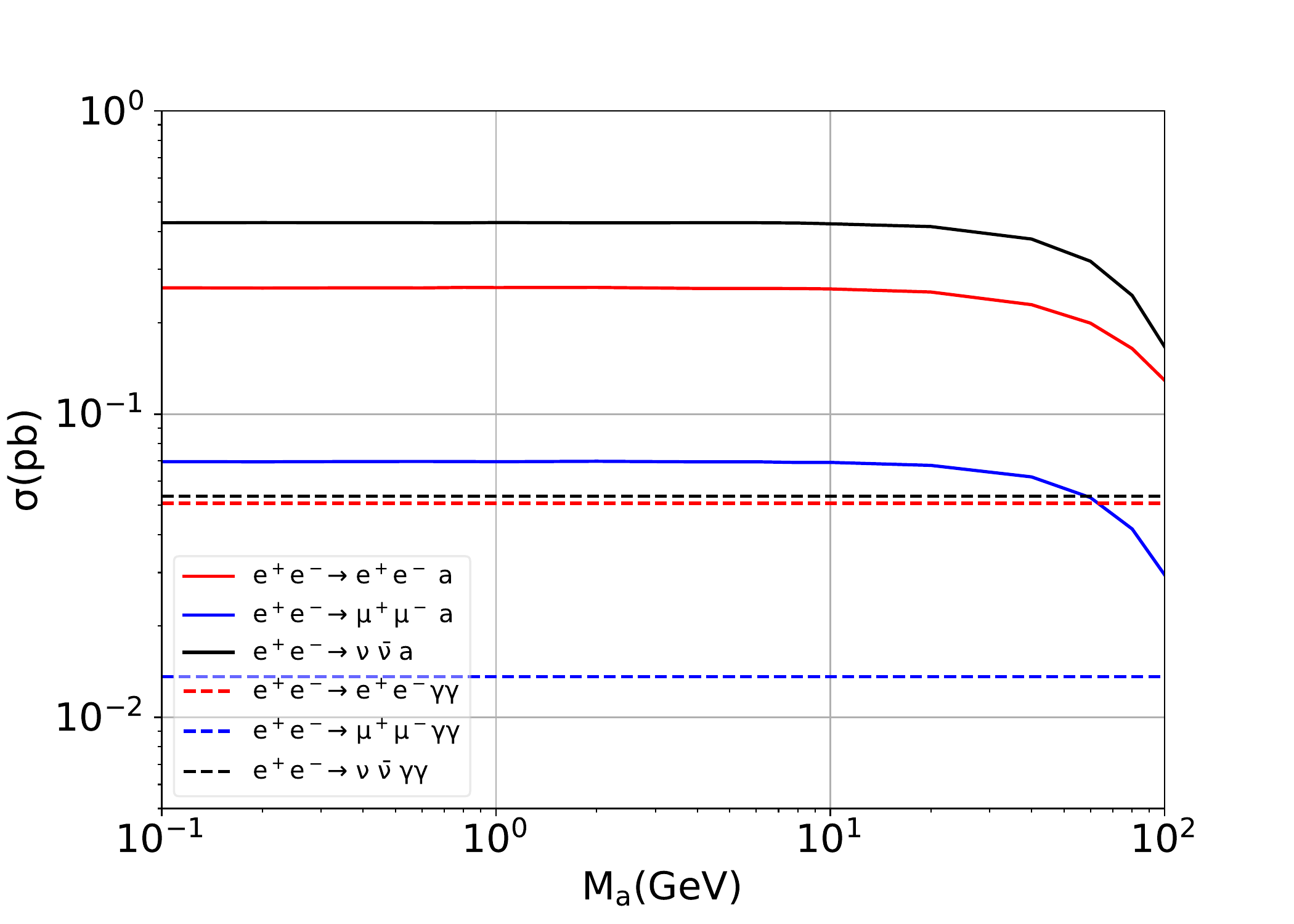}
	\caption{
 \small\label{fig.2} 
 The ALP signal and SM cross-sections at the Higgs factory with $\sqrt{s}=250~\rm GeV$. Signal cross-sections are calculated with the 
 coupling strengths listed in Table.~\ref{Tb.1}: 
 $g_{a\gamma\gamma}=4.88\times10^{-3}~\rm GeV^{-1}$, 
 $g_{aZ\gamma}=1.38\times10^{-3}~\rm GeV^{-1}$, 
 $g_{aZZ}=7.11\times10^{-3}~\rm GeV^{-1}$, and 
 $g_{aWW}=8\times10^{-3}~\rm GeV^{-1}$). 
 }
\end{figure}
\par 

We generated $10^5$ events using $\rm MadGraph5aMC@NLO$. The 
scattering cross-section associated with the process 
$e^+e^-\to f\bar{f}a$ is presented in Fig.\ref{fig.2}, 
where $f=e,\mu,\nu$. 
We have computed the cross-sections using the coupling strengths 
listed in Table~\ref{Tb.1}. Among the three signal processes, 
$e^+e^-\to \nu\bar{\nu}a$ has the largest cross-sections,
as it consists of three flavors of neutrinos. 
On the other hand, $e^+e^-\to\mu^+\mu^-a$ gives the smallest
cross sections.
For $M_a$ ranging from 0.1 GeV to 10 GeV, the cross-section curves 
remain flat. As $M_a$ increases from 10 GeV, the cross sections 
gradually decrease, because the final state phase space becomes limited
with increasing ALP mass. 
This pattern is consistent across all three channels.

\begin{figure}[th!]
	\includegraphics[width=16cm,height=16cm]{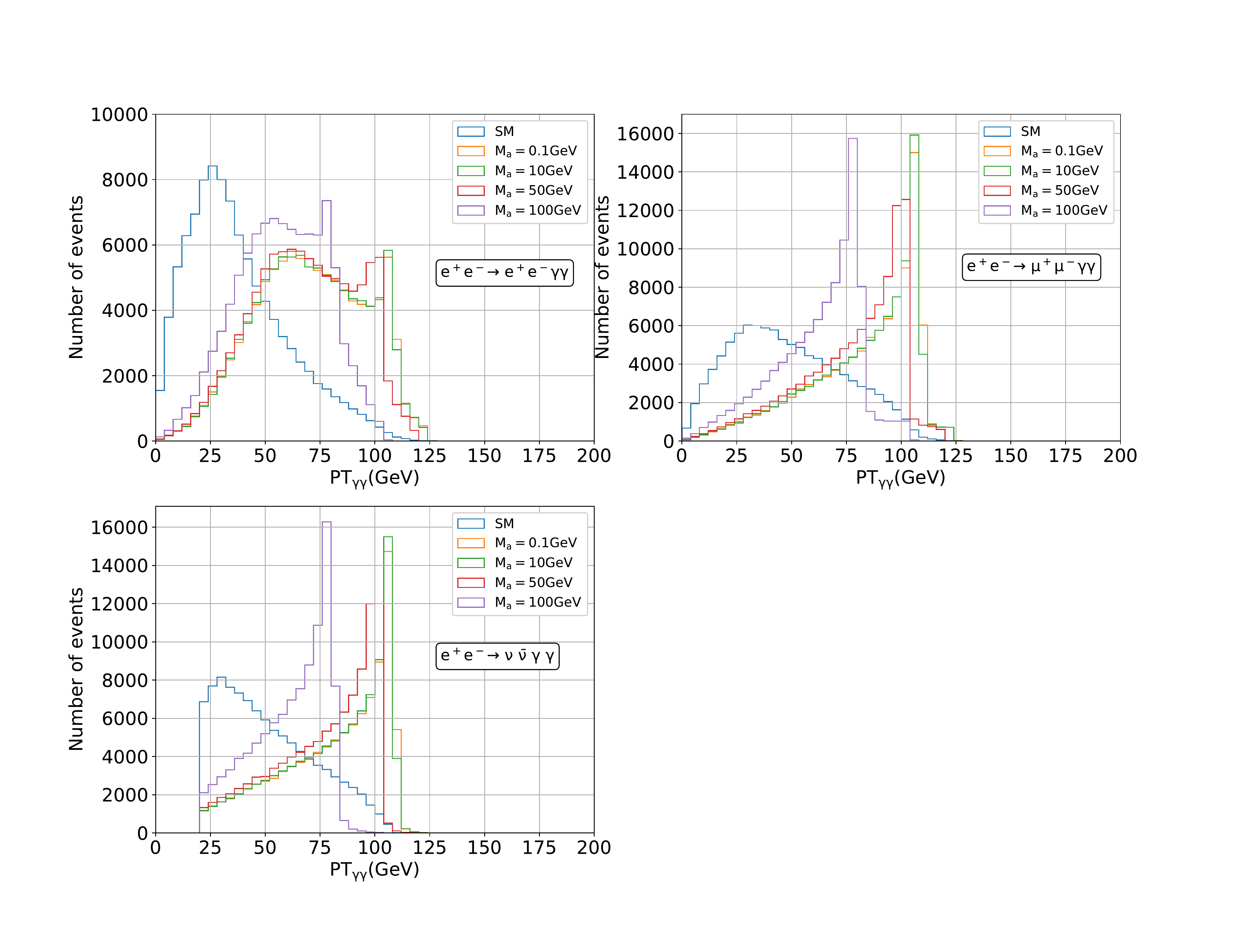}
	\caption{ \small\label{fig.3} 
 Transverse momentum $p_{T_{\gamma\gamma}}$ distributions of 
 the photon pair for the signal processes with 
 $M_a = 0.1 - 100$ GeV  and the corresponding SM background 
 at $e^+ e^-$ colliders with $\sqrt{s}=250~\rm GeV$.}
\end{figure}

To suppress the irreducible background, we apply a cut on the 
transverse momentum of the photon pair.
In Fig.~\ref{fig.3}, we compare the transverse momentum of the 
photon pair for $M_a = 0.1 - 100$ GeV with the corresponding background.
A selection cut of $p_{T_{\gamma\gamma}}>50~\rm GeV$ can suppress the
SM background.

\section{Sensitivity on the ALP Model}
The number of signal events $N_T$ at $e^+ e^-$ colliders
with $\sqrt{s}=250 ~\rm GeV$ is estimated as
\begin{equation}
\label{Eq.8}
N_{T}=\sigma(e^+e^-\to~f\bar{f}~a)\times B(a\to~\gamma\gamma)\times\frac{N(p_{T_{\gamma\gamma}}>50~\rm GeV)}{N_{\rm sim}}\times\mathcal{L} \;,
\end{equation}
where $\sigma(e^+e^-\to~f\bar{f}~a)$ is the ALP production 
cross-section, $B(a\to~\gamma\gamma)$ is the branching ratio of
the ALP to a pair of photons (see Appendix), 
$N(p_{T_{\gamma\gamma}} > 50\, \rm GeV)$ 
is the number of events surviving the $p_{T_{\gamma\gamma}}>$ 50 GeV cut,
and $N_{\rm sim}$ is the total number of events simulated. 
In this study, we generated $N_{\rm sim}= 10^5$ events using 
$\rm MadGraph5 aMC@NLO$ and $\mathcal{L}$ is the integrated luminosity, which we conservatively choose $\mathcal{L}= 2~ab^{-1}$.
Similarly, 
the number of background events $N^{\rm SM}_T$ is estimated as 
\begin{equation}
\label{Eq.9}
   N^{\rm SM}_T=\sigma(e^+e^-\to~f\bar{f}~\gamma\gamma)
   \times\frac{N(p_{T_{\gamma\gamma}}>50~\rm GeV)}{N_{sim}} 
   \times\mathcal{L} \;.
\end{equation}

The number of signal events $N_T$ is proportional to the square of the
ALP coupling strength $g$.  In this study, we consider all possible ALP
interactions encoded in Eq.~(\ref{Eq.3}), from all possible channels 
of ALP
production listed in Fig. \ref{fig.1}. The bound on the ALP coupling as 
a function of ALP mass can be obtained by requiring the 
significance $Z>2$. 
The significance $Z$ is defined as \cite{Arhrib:2019ywg}:
\begin{equation}\label{Eq.10}
Z=\sqrt{2~.\Big[(s+b) \,\ln\Big(\frac{(s+b)(b+\sigma^2_b)}{b^2+(s+b)\sigma^2_b}\Big)-\frac{b^2}{\sigma^2_b}\,ln\Big(1+\frac{\sigma^2_b s}{b(b+\sigma^2_b)}\Big)           \Big]     }\,,
\end{equation}

where the numbers of signal and background events are 
represented by $s$ and $b$, 
respectively. The systematic uncertainty associated with the SM background
$b$ is denoted by $\sigma_b$. A significance value of $Z=2$ is considered,
which corresponds to $95\%$ confidence level (C.L.). 
In the following subsections, we discuss the sensitivity of 
the ALP couplings from ALP production with three different 
leptonic final states at the Higgs factory.

\subsection{ $ e^+ e^- \to e^+ e^- a,\;  a \to \gamma\gamma $}

\begin{figure}[th!]
	\includegraphics[width=16cm,height=8cm]{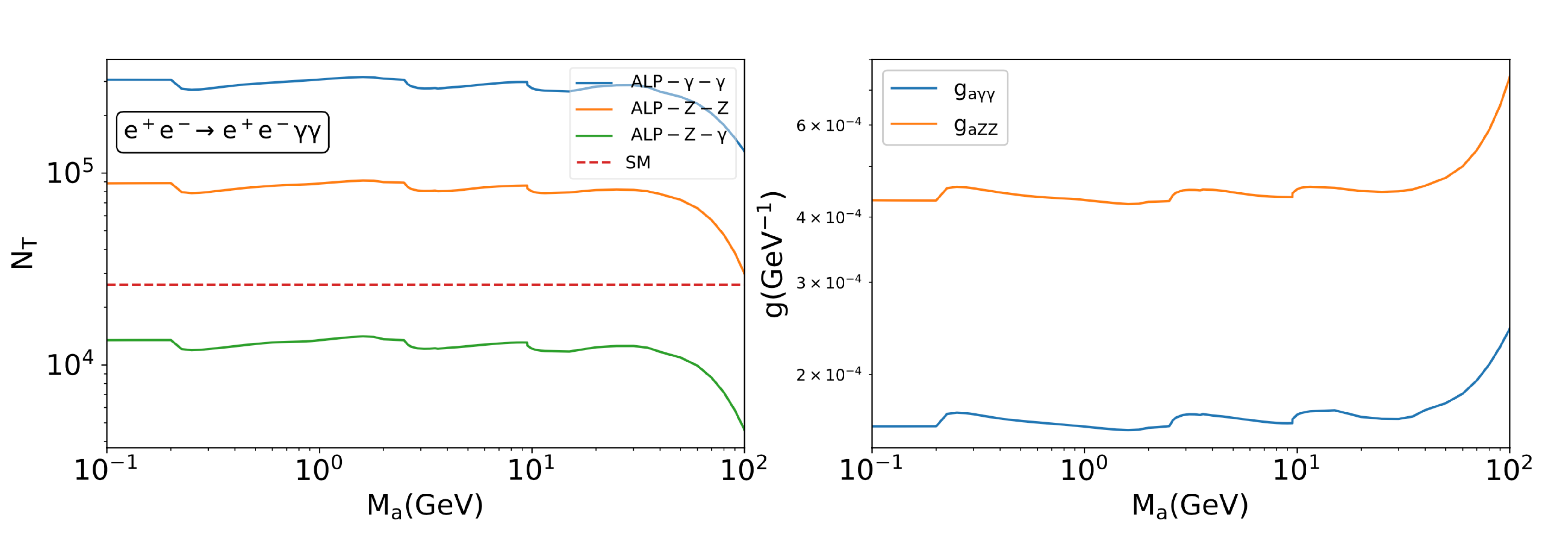}
	\caption{ \small\label{fig.4} 
 Left panel: The number of ALP events from the channel $e^+e^-\to~e^+e^-a$ followed by $a \to \gamma\gamma$ 
 (event rates are estimated with the coupling
 strengths listed in Table.~\ref{Tb.1}: 
 $g_{a\gamma\gamma}=4.88\times10^{-3}~\rm GeV^{-1}$, 
 $g_{aZ\gamma}=1.38\times10^{-3}~\rm GeV^{-1}$, and 
 $g_{aZZ}=7.11\times10^{-3}~\rm GeV^{-1}$].   
 Right panel: The 95$\%$ C.L. sensitivity curves on $g_{a\gamma\gamma}$ 
 (solid blue) and $g_{aZZ}$ (solid orange) for 
 $e^+e^-\to~e^+e^-a,~~a \to \gamma\gamma$.}
\end{figure}

The process of ALP production, in conjunction with a pair of electrons
mediated by $\gamma$ and $Z$ bosons, is illustrated in Fig.\ref{fig.1}. 
In this process, the effective couplings of the ALP to 
$ZZ$, $\gamma\gamma$, and $\gamma Z$ are associated with the
dimensional couplings $g_{aZZ}$, $g_{a\gamma\gamma}$, 
and $g_{aZ\gamma}$, respectively. 
The numbers of signal and background events are estimated using
Eqs.~(\ref{Eq.8}) and (\ref{Eq.9}), and are shown in the left panel of 
Fig.~\ref{fig.4}. 

The combination of production via photon fusion followed by the
ALP decay into diphoton yields the highest number of signal 
events for the specified value of $g_{a\gamma\gamma}$ coupling
listed in Table. \ref{Tb.1}. 
The number of ALP events from the ALP-$ZZ$ vertex is intermediate,
while the ALP-$Z\gamma$ vertex gives the least number of events, 
even the SM event rate is higher than the latter one.
The kinks in the number of signal event curves arise from the 
branching ratio of the ALP into diphoton $a\to\gamma\gamma$.

We then estimate the sensitivity in the ALP couplings versus the
ALP mass, especially $g_{a\gamma\gamma}$ and $g_{aZZ}$ using
Eq.~(\ref{Eq.10}). We account for the systematic uncertainty 
associated with the background estimation by including 
assuming an uncertainty of $\sigma_b=10\%$. The bounds on the 
ALP couplings $g_{a\gamma\gamma}$ (blue) and $g_{aZZ}$ (orange) 
as a function of the
ALP mass $M_a$ are shown in the right panel of Fig.~\ref{fig.4}. 
It is easy to see that the sensitivity of the $g_{a\gamma\gamma}$ 
coupling is a few times better than the $g_{aZZ}$ coupling. 
At lighter ALP mass of $M_{a}=0.1$ GeV, 
the sensitivity of $g_{a\gamma\gamma}$ can reach down to
$\sim 1.5\times10^{-4}\;\rm GeV^{-1}$, 
while $g_{aZZ}$ reaches down to $\sim 4.3\times10^{-4}\;\rm GeV^{-1}$. 
The sensitivity curves stay more or less flat until $M_a = 30\,\rm GeV$ with some irregularities due to the branching ratio into diphoton. As 
$M_a$ increases beyond 30 GeV, the sensitivity is largely worsened due to
smaller phase space for the production of heavier ALPs. 
At $M_a=100$ GeV, the bounds on $g_{a\gamma\gamma}$ and $g_{aZZ}$ are reduced to approximately 
$2.5\times10^{-4}\; \rm GeV^{-1}$ and $7.5\times10^{-4}\;\rm GeV^{-1}$, respectively.

\subsection{$e^+e^- \to \mu^+\mu^- a,~~a \to  \gamma\gamma $}

\begin{figure}[th!]
	\includegraphics[width=16cm,height=8cm]{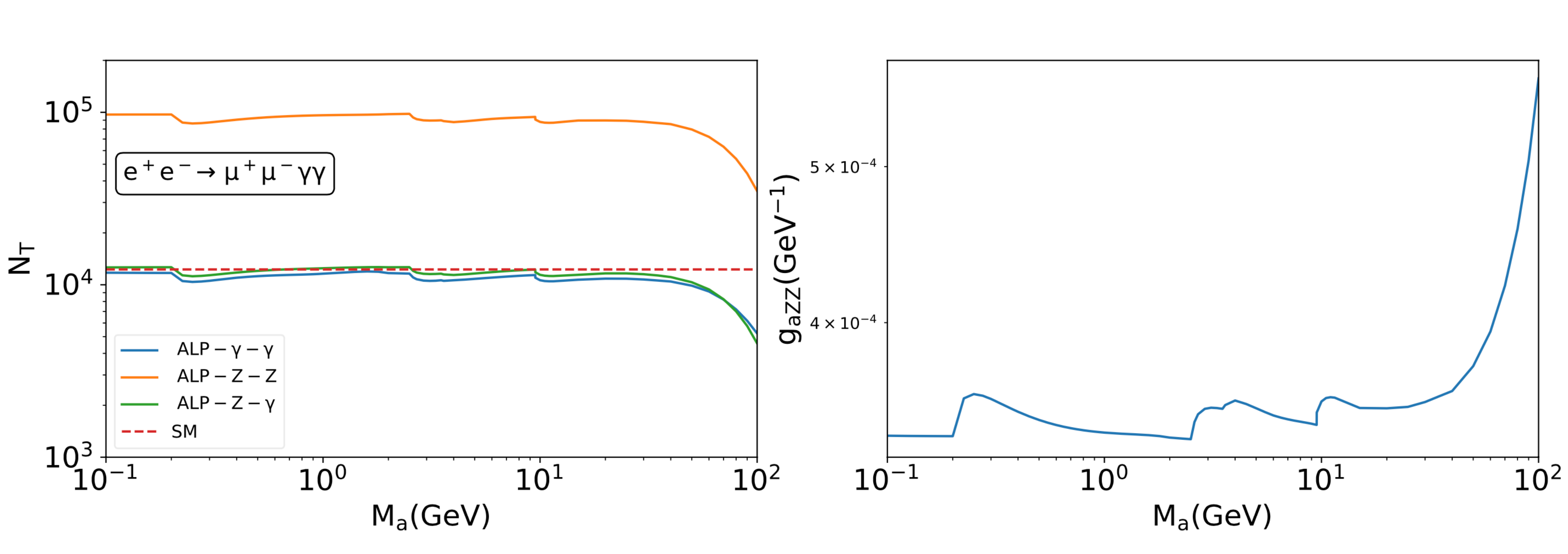}
 \caption{ \small\label{fig.5}  
 Left panel: 
 The number of ALP events from the channel 
 $e^+e^-\to~\mu^+\mu^-a$ followed by $a \to \gamma\gamma$ 
 (event rates are estimated with the 
 coupling strengths listed in Table.~\ref{Tb.1}: 
 $g_{a\gamma\gamma}=4.88\times10^{-3}~\rm GeV^{-1}$, 
 $g_{aZ\gamma}=1.38\times10^{-3}~\rm GeV^{-1}$, and 
 $g_{aZZ}=7.11\times10^{-3}~\rm GeV^{-1}$]. 
 Right panel: The 95$\%$ C.L. sensitivity curve of $g_{aZZ}$ (solid blue) 
 for $e^+e^-\to~\mu^+\mu^-a,~~a \to \gamma\gamma$.}
\end{figure}

Here we consider the associated production of the ALP with 
a $\mu^+\mu^-$ pair. This process only arises from 
$s$-channel diagrams listed in Fig.~\ref{fig.1}.
The left panel of Fig.~\ref{fig.5} shows the number of ALP events 
arising from various ALP vertices.
The highest number of ALP events comes from ALP production associated with the ALP-$ZZ$ vertex. The numbers of ALP events produced via
the ALP-$Z\gamma$ and ALP-$\gamma\gamma$ vertices arre lower than 
that of the SM. 
The right panel of Fig.\ref{fig.5} shows the sensitivty reach
of the $g_{aZZ}$ coupling as a function of ALP mass $M_a$. The 
effect of the diphoton branching ratio also reflects in the sensitivity
curves. At $M_a=0.1~\rm GeV$, $g_{aZZ}$ can be probed down to 
$\sim 3.4\times10^{-4}~\rm GeV^{-1}$. 
The sensitivity of $g_{aZZ}$ weakens with the increment of the ALP mass,
especially for $M_a$ above 30 GeV.

Comparing the bounds of $g_{aZZ}$ obtained in the channels
$e^+e^-\to \mu^+\mu^-a$ (Fig.~\ref{fig.5}) and
$e^+e^-\to e^+e^-a$ (Fig.~\ref{fig.4}), we can see that
$g_{aZZ}$ from the muon channel performs better than the electron 
channel over the entire ALP mass range. This is simply because 
the background in the muon channel is only a fraction of the
electron channel.

\subsection{$e^+e^- \to \nu\bar{\nu}a,\;\; a \to \gamma\gamma $}

\begin{figure}[th!]
	\includegraphics[width=16cm,height=8cm]{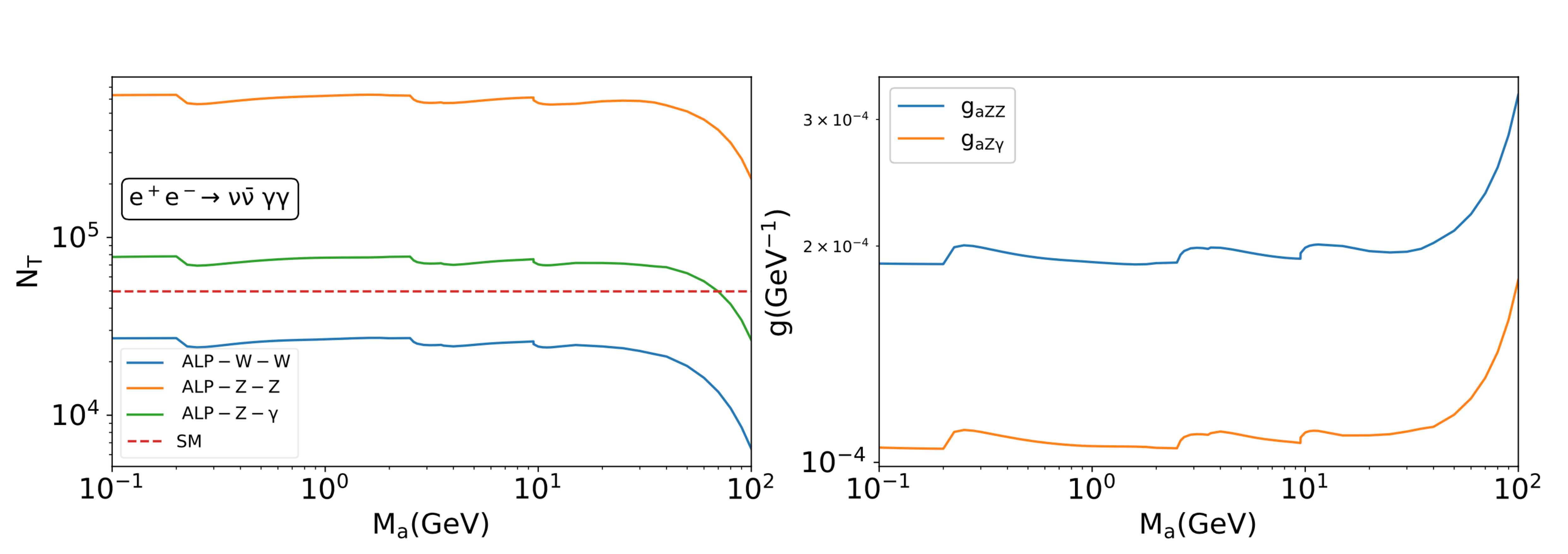}
	\caption{ \small\label{fig.6}  
 Left panel: The number of ALP events from 
 $e^+e^-\to~\nu~\bar{\nu}~a$, followed by $a \to \gamma\gamma$ 
 (event rates are estimated with the 
 coupling strengths listed in 
 Table.~\ref{Tb.1}: $g_{aZ\gamma}=1.38\times10^{-3}~\rm GeV^{-1}$, 
 $g_{aZZ}=7.11\times10^{-3}~\rm GeV^{-1}$, and 
 $g_{aWW}=8\times10^{-3}~\rm GeV^{-1}$]. 
 Right panel: The 95$\%$ C.L. sensitivity curves on $g_{aZZ}$ (solid blue) and $g_{aZ\gamma}$ (solid orange) for 
 $e^+e^-\to~\nu~\bar{\nu}~a,~~a \to \gamma\gamma$.}
\end{figure}

As already shown in Fig.~\ref{fig.2}, the channel 
$e^+e^- \to \nu \bar{\nu} a$ with $a \to  \gamma\gamma$ 
has the largest cross-sections compared to the other two processes.
This process also presents an opportunity to investigate the 
ALP-$WW$ vertex. In addition to the ALP-$WW$ vertex,
the ALP-$ZZ$ and ALP-$Z\gamma$ vertices also make contributions,
which are depicted in Fig.~\ref{fig.1}. 

The number of ALP events from the ALP-$ZZ$ vertex is higher than 
that from the other two vertices. The ALP production rate from the 
ALP-$WW$ vertex is the lowest and is even lower than that of the SM. 

The bounds on $g_{aZZ}$ and $g_{aZ\gamma}$ couplings are shown in 
the right panel of Fig.\ref{fig.6}. In this case, the $g_{aZ\gamma}$ coupling has a better bound compared to the $g_{aZZ}$ coupling. At $M_a$=0.1 GeV, the $g_{aZ\gamma}$ coupling can reach down to
$\sim 10^{-4}\;\rm GeV^{-1}$, while the $g_{aZZ}$ coupling 
reaches down to $1.8\times10^{-4}\; ~\rm GeV^{-1}$. 
Similar to previous cases, the sensitivity of the couplings weakens 
as the ALP mass $M_a$ increases.

When comparing the bounds of $g_{aZZ}$ coupling 
obtained from all different channels 
the best sensitivity comes from 
$e^+e^- \to \nu\bar{\nu}a\,, \;\; a \to \gamma\gamma$.
At $M_a$=0.1\;\rm  GeV, the $g_{aZZ}$ coupling reaches down to
$1.8\times10^{-4}\rm GeV^{-1}$.
The $e^+e^- \to e^+e^- a,\; ~~a \to \gamma\gamma$ channel offers 
the least sensitivity (for $M_a$=0.1 GeV $g_{aZZ}$ coupling  
only reaches down to $4.3\times10^{-4}\rm GeV^{-1}$). 
The limit from the $e^+e^- \to \mu^+\mu^- a,\;\; ~~a \to \gamma\gamma$ 
channel is intermediate (for $M_a$=0.1 GeV the $g_{aZZ}$ coupling reaches 
down to $3.4\times10^{-4}\rm GeV^{-1}$). 
This trend is visible across the entire ALP mass range from $M_a$ = 0.1 GeV to 100 GeV.

\begin{figure}[th!]
	\includegraphics[width=14cm,height=10cm]{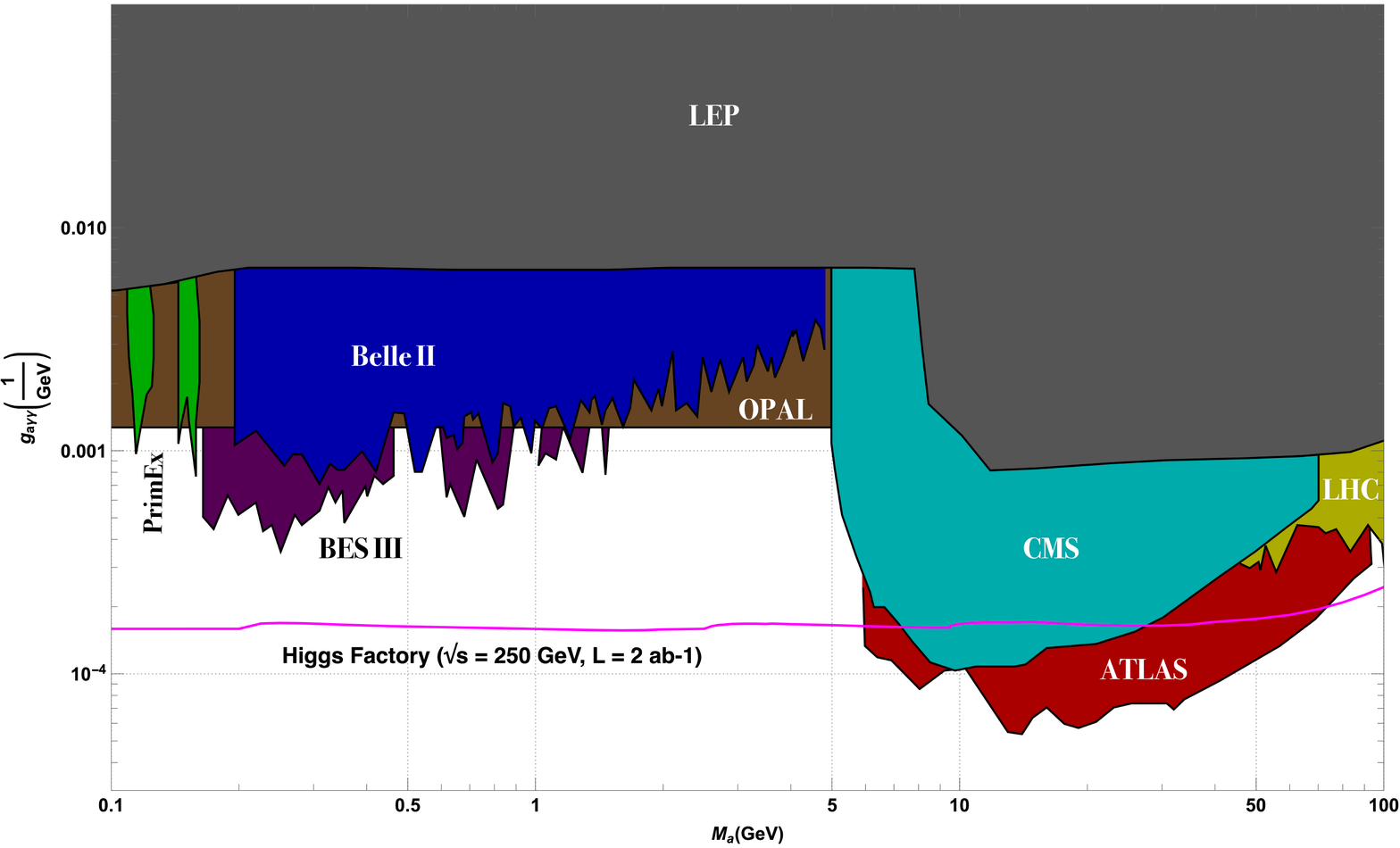}
	\caption{\small\label{fig.7}
Summary plot of the sensitivity of $g_{a\gamma\gamma}$ that we can
achieve at the Higgs factory $\sqrt{s}=250$ GeV with an integrated
luminosity 2 ab$^{-1}$, and compared with other existing constraints. Existing constraints in the figure include PrimEx~\cite{Aloni:2019ruo}, BES III~\cite{BESIII:2022rzz}, Belle II\cite{Belle-II:2020jti}, LEP~\cite{Jaeckel:2015jla}, OPAL~\cite{Knapen:2016moh}, CMS~\cite{CMS:2018erd}, ATLAS~\cite{ATLAS:2020hii} and LHC~\cite{Knapen:2016moh} (extracted from the GitHub page~ \cite{AxionLimits}).
 }
 \end{figure}

\section{Conclusions}

In this study, we have explored the sensitivity potential of the 
future Higgs factories, including ILC, CEPC, and FCC-ee, on 
probing dimensionful coupling constants $g_{a\gamma\gamma}$,
$g_{aZ\gamma}$, $g_{aWW}$, and $g_{aZZ}$ of the axion-like particle, via
the processes $e^+ e^- \to f \bar f a \; (f=e, \mu,\nu)$ 
followed by $a\to \gamma\gamma$. We used a center-of-mass energy 
$\sqrt{s}=250$ GeV with an integrated luminosity 2 ab$^{-1}$.

Our results have shown that the channel $e^+e^-\to e^+e^-a, \;\; 
a \to \gamma\gamma$ provides the best bound for 
the $g_{a\gamma\gamma}$ coupling, 
while the process $e^+e^-\to \nu\bar{\nu}a,\; \; 
a \to \gamma\gamma$ offers the best bound for the $g_{aZZ}$ and 
$g_{aZ\gamma}$ couplings. 

Without loss of generality, we have used $C_{WW}=2$ and $C_{BB}=1$ 
such that $g_{a\gamma\gamma}$, $g_{aZ\gamma}$, $g_{aWW}$, and $g_{aZZ}$
are related to one another shown in Eqs.~(\ref{Eq.4}) -- (\ref{Eq.7}),
and they are all nonzero. 
We can easily extend the analysis to independent coupling strengths.

Finally, we show in Fig.~\ref{fig.7} the summary plot of the 
sensitivity of $g_{a\gamma\gamma}$ that we can achieve at the Higgs 
factories, and compared with other existing constraints. The sensitivity
can improve down to about $1.5\times 10^{-4}\; {\rm GeV}^{-1}$ 
over the mass range of $M_a = 0.1-6$~GeV, as well as a small corner
at $M_a \simeq 70-100$~GeV.

Our estimates of the bounds for the $g_{a\gamma\gamma}$, $g_{aZ\gamma}$,
and $g_{aZZ}$ couplings as a function of ALP mass ($M_a$) ranging from 
0.1 GeV to 100 GeV provide valuable insights for future experiments 
aiming to detect ALPs.

\appendix

\section{Partial Decay Widths and Branching Ratios of the ALP}
\label{app}
The two-body partial decay widths of the ALP to photons and fermions
are given below. The branching ratios are evaluated with 
$C_{BB}=1$, $C_{WW}=2$, $c_{a\phi}=1$ and $f_a=1000~GeV$. 
Here $M_l$ and $M_q$ are the masses of charged leptons and quarks.
\begin{eqnarray}
\Gamma(a\to \gamma\gamma) &=& 
  \frac{M_{a}^6(C_{BB}c_w^2+C_{WW}s^2_w)^2}{4f_a^2\pi|M_a^3|}\;,\\
\Gamma(a\to l\bar{l}) &=& \frac{c^2_{a\phi}M_a^2\sqrt{M_a^2-4M_a^2M_l^2}~vev^2~y_l^2}{16\pi f_a^2|M_a^3|}
    \;\\
\Gamma(a\to q\bar{q}) &=& \frac{3~c^2_{a\phi}M_a^2\sqrt{M_a^2-4M_a^2M_q^2}~vev^2~y_q^2}{16\pi f_a^2|M_a^3|}
    \;
\end{eqnarray}

\begin{figure}[th!]
	\includegraphics[width=15cm,height=8cm]{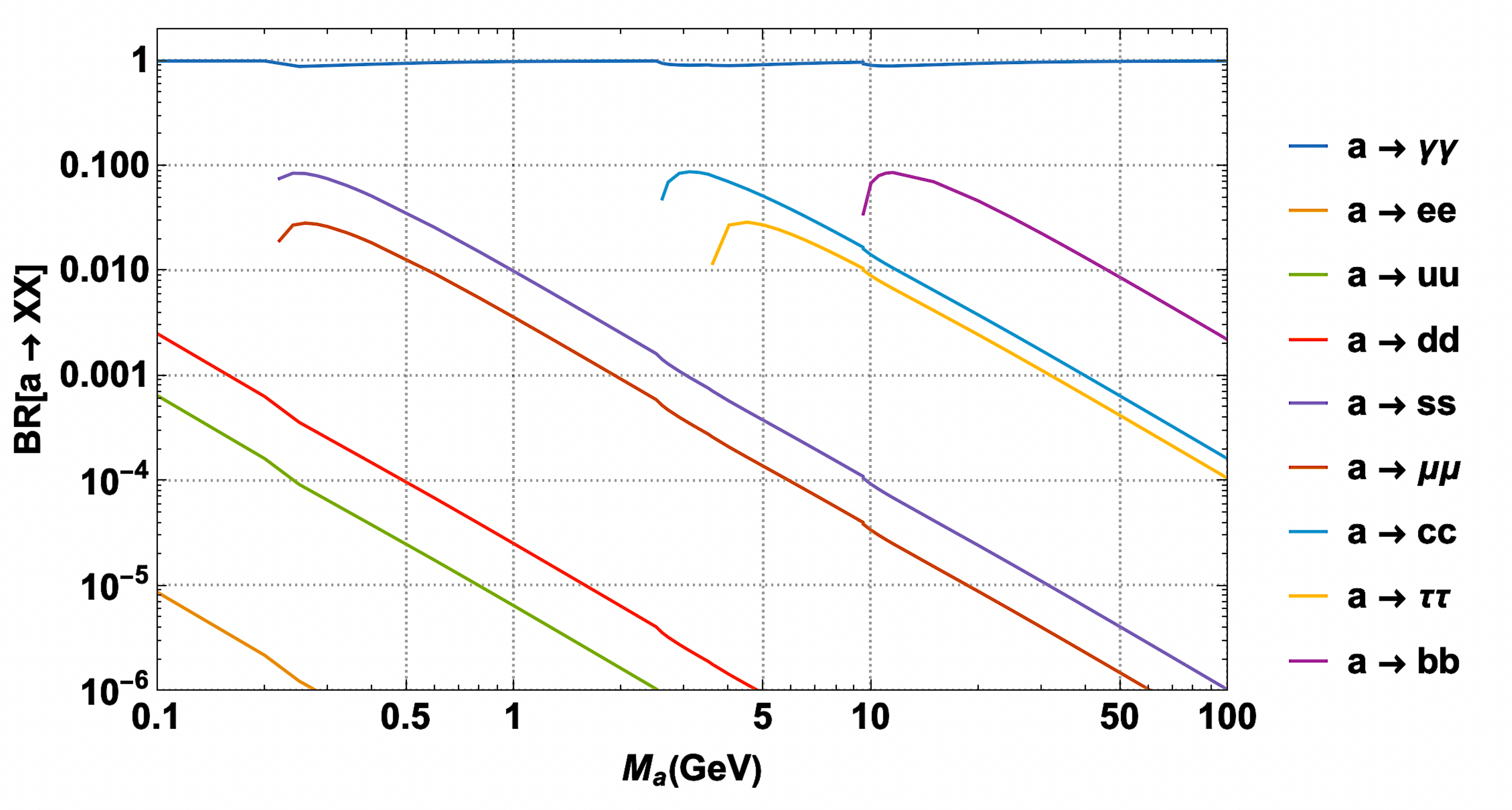}
	\caption{ \small\label{fig.8} Branching ratios of the ALP
 with $C_{WW}=2$, $C_{BB}=1$, and $f_a=1$~TeV.
 }
\end{figure}

\section*{Acknowledgement}
Special thanks to Zeren Simon Wang and Nguyen Tran Quang Thong for an enlightening discussion.
The work was supported in part by NSTC under the grant number
MOST-110-2112-M-007-017-MY3.

\bibliographystyle{JHEP.bst}
\bibliography{biblio.bib}
\end{document}